\documentclass[12pt]{article}
\usepackage{epsfig,rotating,subeqn}
\usepackage{graphicx} 
\usepackage{bm}
\usepackage{color}
\newcommand{\beq}{\begin{equation}}
\newcommand{\eeq}{\end{equation}}
\newcommand{\ba}{\begin{array}}
\newcommand{\ea}{\end{array}}
\newcommand{\lsim}   {\mathrel{\mathop{\kern 0pt \rlap
  {\raise.2ex\hbox{$<$}}}
  \lower.9ex\hbox{\kern-.190em $\sim$}}}
\newcommand{\gsim}   {\mathrel{\mathop{\kern 0pt \rlap
  {\raise.2ex\hbox{$>$}}}
\lower.9ex\hbox{\kern-.190em $\sim$}}}

\begin{document}
\title{{\bf Ultra High Energy Cosmic Rays: \\ 
               Anisotropies and Spectrum}}
     
\author{R. Aloisio$^{1}$ and D. Boncioli$^{2}$\\
        {\it\small $^1$INFN, Laboratori Nazionali del Gran Sasso, Assergi (AQ), Italy} \\
        {\it\small $^2$Universit\`a di Roma II Tor Vergata and INFN, Roma, Italy}        }

\maketitle

\bigskip

\abstract{
The recent results of the Pierre Auger Observatory on the possible correlation of Ultra High Energy Cosmic Rays 
events and several nearby discrete sources could be the starting point of a new era with charged particles 
astronomy. In this paper we introduce a simple model to determine the effects of any local distribution 
of sources on the expected flux. We consider two populations of sources: faraway sources uniformly 
distributed and local point sources. We study the effects on the expected flux of the local distribution
of sources, referring also to the set of astrophysical objects whose correlation with the Auger 
events is experimentally claimed. 
}

\section{Introduction}
\label{introduction}

Ultra High Energy Cosmic Rays (UHECR) are the most energetic particles known in nature, 
with observed energies larger than $10^{18}$ eV. The detection of these particles, started already 
in the 50s with the pioneering experiments of Volcano Ranch in the USA and the Moscow University array
in the USSR, poses many interesting questions mainly on their origin and chemical composition. 
In the recent years a new step forward in unveiling the nature of UHECR was done with the measurements
performed by HiRes and AGASA first, and nowadays with the first results of the Pierre Auger
Observatory in Argentina.  

Soon after the discovery of the first UHECR event with energy around $10^{20}$ eV \cite{Linsley63} 
it becomes clear that these extreme high energy particles could reach us only from the nearby universe; 
in other words, at the highest energies the universe contributing to the observed flux is definitively confined 
inside about 200 Mpc around us. In particular, if UHECR are mainly composed by protons, the interaction with the
Cosmic Microwave Background (CMB) which produces a photo-pion production process, leads to the well 
known GZK suppression in the flux \cite{GZK}. At energies around $5\times 10^{19}$ eV the visible universe in
protons rapidly passes from Gpc scale to Mpc scale. On the other hand, if UHECR are mainly composed by
nuclei, the photo-disintegration process suffered on astrophysical backgrounds (not only the CMB but also the
Infrared/Optical Background (IRB)) causes a strong depletion in the observed spectrum; this depletion occurs at
energies that span from $2\times 10^{19}$ eV (for the lightest nuclei) up to $2\times 10^{20}$ in the case of Iron
\cite{NucleiCut, NucleiNoi}. At these energies, as for protons, the contributing universe rapidly decreases 
from Gpc to Mpc scale. 

In the case in which the relatively nearby sources of UHECR are not uniformly distributed, as in the case
of astrophysical sources, the arrival directions of the most energetics UHECR could reflect such 
anisotropy. This expectation depends mainly on the electrical charge of UHECR because of the effect 
of the intervening magnetic fields, in particular of the galactic magnetic field. In the case of protons the typical 
deflection angle on a $\mu$-Gauss scale magnetic field and over Kpc distances is of the order of few 
degrees for particles with energy $E>50$ EeV. With such angular resolution, inside distances at the 
Mpc scale, it is possible to resolve many interesting astrophysical objects that, according to several models, 
could be the sources of UHECR. Increasing the electric charge at $E>50$ EeV, already in the case of Helium 
(Z=2), the deflection angle becomes larger than 10 degrees, arriving up to more than 50 degrees in
the case of Iron (Z=26). Therefore, if UHECR are mainly composed by nuclei it will be nearly impossible to
observe any correlation with sources. 

Let us now discuss the experimental situation concerning signals of anisotropy in the arrival directions 
of UHECR. There are several experimental results that, in the past, claimed signals of anisotropy.
In the energy range $0.1~ < E < ~10$ EeV an excess of events from the Galactic Center was reported by
SUGAR and AGASA experiments \cite{SUGAR-AGASA}. This possibility was investigated also by the Pierre
Auger collaboration that, with the data collected till March 2007, claimed a negative result that did not confirm the
SUGAR and AGASA excess \cite{Auger_GC1}. On the other hand, going to higher energies where the 
contributing universe is sensibly reduced, the AGASA collaboration searched for clusterings of events in the
Northern sky founding a triplet whose chance probability was estimated to be of the order of 1\% 
\cite{AGASA_3}. This triplet was found to be correlated with a HIRES high energy event \cite{HiRes_3}. The
Pierre Auger collaboration found a deviation from isotropy for events above 52 EeV with a chance probability of 
less than 1\% \cite{Auger_ani}. 

Another way of looking at anisotropy is to search for correlations in the direction of known astrophysical 
objects that could be the sources of UHECR. In this case, even if there is no excess over the expected 
background, the event sample could anyway exhibit a correlation with the direction of sources fixed 
a priori. In the past, searches of this kind were performed using a collection of data by different experiments 
and some hints of correlation were identified with a particular kind of AGN: the BL Lacertae objects (BL-Lacs),
AGN with the jet pointed toward us \cite{BL-Lacs}. This result was recently questioned by the Auger
collaboration that performing the same search obtained a negative result \cite{Auger-BL-Lacs}; it has 
to be pointed that the Auger observations refer to the southern hemisphere while all the other to the northern
one. 

Apart from the specific case of BL-Lacs, in general, there have been many speculations on whether AGNs
are the actual sources of UHECR \cite{AGN}. In this case the brightest and closest AGN could produce the
UHECR observed on earth and the arrival directions of such particles could point back to their source. The large
number of identified AGNs makes these objects a good candidate for studying possible correlations with UHECR.

The Auger collaboration performed a search for correlation of their events with AGNs from the 12th edition of the
catalog of quasars and active galactic nuclei by V\'eron-Cetty and V\'eron \cite{VCVcat} (VCV catalog). 
The correlation study is based on three main parameters: the maximum difference in angle between the UHECR
arrival direction and the AGN direction $\psi_{0}$, the minimum energy of cosmic rays showing the correlation
$E_{0}$ and the maximum red-shift of the correlated AGNs $z_{0}$. 
Using the data collected between January 1 2004 and August 31 2007 the collaboration found that 20 out 
of 27 events correlated with at least one of the selected AGNs \cite{AugerCorrSc,AugerCorrApp}. In this 
analysis the parameters that minimize the probability of having a chance correlation with isotropic events 
were found to be $\psi_{0}=3.2^{\circ}$,  $z_{0}=0.017$, $E_{0}=57$ EeV.  An updated analysis of such
correlation was performed with the cosmic rays events collected up to March 31 2009: 17 out of 44 
independent events were found to correlate, within roughly the same choice of parameters of 2007. This
updated analysis neither strengthens the case for anisotropy, nor does contradict the earlier results
\cite{Auger_corr_up}.

The HIRES collaborations performed the same analysis of the Auger collaboration for correlations between
stereo events and AGNs from the VCV catalog but no significant correlation was found
\cite{Hires_agn}.

In the present paper, using the result of the Auger collaboration on correlations, we study the expected UHECR
spectrum assuming that UHECR are mainly composed by protons. The dip model, proposed already in 2002 
\cite{dip}, is based on this assumption and explains the behavior of the observed spectrum by the pair-production
energy losses suffered by protons on the CMB radiation field. Depending on the injection of protons at the
source, the process of pair production on the CMB photons produces a dip in the expected spectrum on earth;
this dip, placed in the energy range between $2\times 10^{18}$ and  $4\times 10^{19}$ eV, is observed by
different experiments: Akeno-AGASA, HiRes, Yakutsk \cite{dip}.

The dip model was also studied in the framework of the results of the Auger collaboration; in \cite{dip} this
 comparison showed a
good agreement of the Auger data of 2007 with the pair production dip. The new release of Auger data on the
observed UHECR spectrum shows a steepening of the spectrum at the highest energies not much
consistent with the predicted shape of the GZK cut-off, therefore showing a less significative agreement with a
proton dominated spectrum \cite{disapp}. Moreover, the chemical composition observed by the Auger
collaboration favours a nuclei
dominated spectrum progressively heavier in the energy region (4 - 40) EeV \cite{Auger_Xmax}. These two
evidences have 
recently triggered a new possible explanation of the Auger flux in terms of a two component spectrum: a lighter
(proton dominated) component at energies in the range (0.1 - 1) EeV and an heavier (nuclei dominated)
component at higher energies \cite{disapp}.  Concerning energies below $10^{18}$ eV,
some constraints can be found on the proton flux if the cosmic rays sources are optically thin and emit neutrinos
\cite{neutrinos}.

A nuclei dominated spectrum at the highest energies is hardly compatible with the correlations observed by the
Auger collaboration.
In order to combine correlations and spectrum we consider here the case of a pure proton composition, with
particles injected by two different classes of sources: homogeneously distributed at red-shift $z>z_{0}$
and a set of point sources at $z \leq z_{0}$. Our aim is also to determine possible features in the spectrum 
that can be connected with a particular local distribution of the sources. 

The paper is organized as follows: in section \ref{sec:model} we discuss the model while in section
\ref{sec:spectra} we compute the UHECR fluxes, studying also different choices for the contributing local
sources; a discussion of the results and the conclusions take place in section \ref{sec:conclu}.

\section{The model}
\label{sec:model}

The approach used in this paper is based on the hypothesis of continuous energy losses (CEL) of the 
propagating protons. In the propagation through astrophysical backgrounds (mainly CMB) the interaction of
protons is naturally affected by fluctuations, with a non-zero probability for a particle to travel without loosing
energy. In the CEL approximation such fluctuations are neglected; as was shown in
\cite{dip,BereGrigo,BereKach} this approach has a limited effect on the flux computation. Only at the highest
energies ($E>100$ EeV) fluctuations produce a deviation of the order of $10\%$ of the CEL flux respect to the
flux computed with a standard Monte Carlo simulation \cite{BereKach}. Under the CEL hypothesis the evolution
of the proton energy is described by a simple differential equation, whose solution gives the proton energy
$E_g(z)$ at any epoch $z$. One has:
\begin{equation}
\frac{dE_g(z)}{dz}=E_g (1+z)^3 \left | \frac{dt}{dz} \right | \beta_0((1+z)E_g),
\label{eq:dEgdz}
\end{equation}
being $\beta_0$ the rate of energy losses at red-shift zero suffered by protons on the CMB field, due to 
the processes of pair production and photo-pion production, as well as the losses suffered because of 
the adiabatic expansion of the Universe \cite{dip,BereGrigo}. The quantity $dt/dz$ fixes the cosmology: 
\begin{equation}
\left | \frac{dt}{dz}\right | = \frac{1}{H_0 (1+z)\sqrt{\Omega_m (1+z)^3 + \Omega_\Lambda}},
\label{eq:dtdz}
\end{equation}
where our choice for the Hubble constant and the matter/vacuum density are (throughout the whole paper) 
$H_0=70$ km/Mpc/s, $\Omega_m=0.3$, $\Omega_{\Lambda}=0.7$.

Equation (\ref{eq:dEgdz}) can be solved fixing as initial condition the observed energy $E$ on earth (at $z=0$), 
obtaining the function $E_g(E,z)$ (with $E_g(E,z=0)=E$).

As was shown in \cite{dip,BereGrigo,BereKach} the CEL hypothesis is very well suited to compute the
expected flux of UHE protons from both a continuous distribution of sources as well as from a set of sparse 
point sources. In the present study, as discussed in the introduction, driven by the Auger collaboration results on 
anisotropy, we consider two different families of sources: continuous distributed at high red-shifts and 
a set of point sources in the local Universe. 
The red-shift $z_0$ that corresponds to the distance at which the distribution of UHECR events starts to feel
the granularity of the local universe can be considered as a parameter of our building model. In our analysis
we will consider two different possibilities to fix $z_0$. In the first case we will directly refer to the Auger analysis, 
where $z_0$ is assumed to be the maximum red-shift of the AGNs, taken from the VCV
catalog, whose position in the sky can be correlated with the Auger events. In this case, as measured by the Auger
collaboration \cite{Auger_ani}, the value is fixed to $z_0=0.017$. In the second case we will vary $z_0$ 
taking into account its physical role, as the maximum distance after which a distribution of UHECR sources, 
with a given number density, cannot be resolved by the detector in terms of single point sources, 
because of magnetic deflections. At energies larger than $E_0=57$ EeV the protons path length rapidly falls
becoming less than $50$ Mpc at $E>100$ EeV. This fixes the scale of the parameter $z_0$ that should 
correspond to an angular distance not larger than $100\div 300$ Mpc, therefore we will consider different values of
$z_0$ in the range $z_0=0.01\div 0.1$. In this context the choice of $z_0$ roughly corresponds to the so-called 
GZK sphere inside which the propagation of UHE protons is less affected by the photo-pion production process. 

Let us now discuss the injection function associated to the sources. We will assume all sources of each
population identical with injection given by:
\begin{equation}
Q(E_g)=k(\gamma_g) \frac{L}{m_p^2} (1+z)^{m} \left (\frac{E_g}{m_p} \right )^{-\gamma_{g}} ~,
\label{eq:inj}
\end{equation}
being $m_p$ the proton mass, $L$ the source luminosity in UHECR, $\gamma_{g}$ the injection power law index
at the source, $k(\gamma_{g})$ a normalization constant ($k(\gamma_{g})=\gamma_{g} -2$ for 
$\gamma_{g}>2$ and $k(\gamma_{g})=[\ln(E_{max}/m_p)]^{-1}$ for $\gamma_{g}=2$) and $z$ the red-shift.
The term $(1+z)^{m}$ describes the possible cosmological evolution of the source, i.e. the increase of
luminosities and/or space densities with red-shift observed for many astronomical populations. In particular 
AGNs exhibit the evolution seen in radio, optical and X-ray observations (see \cite{dip} and references therein).
Apart from $L$, $\gamma_{g}$ and $m$ another important parameter fixes the source behavior, namely the
maximum energy that a proton can acquire at the source $E_{max}$. As we will discuss in the following this 
parameter is particularly important in fixing the flux behavior. 

The first population of sources we consider here is a set of point sources with a steady isotropic emission
and placed at different red-shifts (inside $z_0$) and along different directions. The flux of UHE protons produced
by a single source at red-shift $z^*$ can be written as \cite{book}:

\begin{equation}
J_{d}(E,z^*)=\frac{1}{(4\pi)}\frac{Q_d(E_g(E,z^*))}{(1+z^*)(R(t_0)r)^2}
\frac{\partial E_g(E,z^*)}{\partial E}
\label{eq:Jdiscrete}
\end{equation}
where $Q_d$ is the injection function of a single local source, $E_g(E,z)$ is the solution of the losses equation
(\ref{eq:dEgdz}) and $(\partial E_g/\partial E)$ is given in \cite{dip,BereGrigo}. Using as before the standard
cosmology, the source distance as function of the red-shift is given by
\begin{equation}
R(t_0)r =\frac{c}{H_0}\int_0^{z^*} \frac{dz}{\sqrt{(1+z^3)\omega_m+\omega_\Lambda}}~.
\end{equation}

The flux of UHE protons produced by a homogeneous distribution of sources can be written as the integral over 
the comoving volume of the flux by a single source \cite{dip}:
\begin{equation}
J_{h}(E)= \frac{1}{(4\pi)^2} \int dV \frac{{\cal Q}_h(E_g(E,z))}{(1+z)(R(t_0)r)^2}\frac{\partial E_g(E,z)}{\partial E} .
\label{eq:Jhomo}
\end{equation}
The quantity ${\cal Q}_h$ represents the number of particles injected per unit 
time, energy and volume by the homogeneous distribution of sources
\begin{equation}
{\cal Q}_h=n_s Q_h(E_g) \Theta(z-z_0)=k(\gamma_g)\frac{{\cal L}_h}{m_p^2}(1+z)^m 
\left (\frac{E}{m_p^2}\right )^{-\gamma_g}  \Theta(z-z_0)
\label{eq:inj_h}
\end{equation}
being ${\cal L}_h=n_s L_h$ the emissivity of the homogeneous distributed sources, i.e. the number of
particles injected (in UHECR) per unit time and volume. As discussed above, in equation (\ref{eq:inj_h}) 
we have added a $\Theta$-term because we assume the homogeneous distribution only at red-shift larger than
$z_0$. The (comoving) volume integral in equation (\ref{eq:Jhomo}) can be easily transformed in a red-shift
integration \cite{dip}, being the maximum red-shift of integration $z_{max}$ the solution
of the equation $E_g(E,z_{max})=E_{max}$. 

Let us point out that the flux (\ref{eq:Jhomo}) from homogeneously distributed sources is characterized
only by the emissivity ${\cal L}_h$ (number of particles emitted per unit time and volume), which entangles two 
physical informations: the luminosity of sources in UHECR and their number density. This fact is a natural
consequence of the hypothesis of unresolved sources at red-shift $z>z_0$. On the other hand, in the case of
local sources, the number of expected events depends only on the source luminosity as follows from equations
(\ref{eq:inj}) and (\ref{eq:Jdiscrete}), being the information on the sources number density already in the local
(resolved) distribution.

As anticipated in the introduction, we are mainly interested in the effects on the UHECR flux of the 
local distribution of sources, therefore we should take into account the number of events expected in a fixed 
direction. To compute it we introduce the relative acceptance $\omega$ of the detector following \cite{Sommers}.
The number of the detected CR events are indeed distributed in the sky depending on both the real celestial
anisotropy and the detector relative exposure $\omega$.

The relative exposure, that depends only on the declination $\delta$, is normalized as: $$2\pi \int 
\omega(\delta) \cos\delta d\delta= \Delta\Omega_{m} ~,$$ being $\Delta\Omega_{m}=\pi \sin^2 \theta_m$ 
and $\theta_m=60^{\circ}$.

In the case of point sources the angular dependence of the expected flux will be also 
affected by the direction associated to the contributing source. Being $\hat{\Omega}^*$ this direction 
we can rewrite the expected flux taking into account its angular dependence as 
$$4\pi J_d(E,z^*) \delta_D(\hat{\Omega}-\hat{\Omega}^*)~,$$
here and in the following $\delta_D$ is the Dirac delta function. The specific direction $\hat{\Omega}^*$
can be expressed in terms of the equatorial terrestrial coordinates $(\alpha,\delta)$ simply using the 
transformation
$$\delta_D(\hat{\Omega}-\hat{\Omega}^*)=\frac{1}{\cos \delta }\delta_D(\alpha-\alpha^*)
\delta_D(\delta-\delta^*) ~.$$

Taking into account the Auger observatory acceptance function and the angular dependence due to point 
sources, we can rewrite the total expected UHECR flux as a function of the equatorial terrestrial coordinates
as: 
\begin{equation}
J_{tot}(E,\alpha,\delta)=\omega(\delta) \left [ J_{h}(E) + 
\sum_i J_{d}(E,z_i) \frac{4\pi}{\cos{(\delta)}} \delta_D(\alpha-\alpha_i)\delta_D(\delta-\delta_i) \right ],
\label{eq:total1}
\end{equation}
being $J_h$ the contribution to the flux due to the homogeneous distributed sources and $J_d$ the 
flux of the local point sources placed at red-shifts $z_i$ and with equatorial terrestrial 
coordinates $(\alpha_i,\delta_i)$. Using equation (\ref{eq:total1}) it is possible to determine the number 
of events at energy $E\ge E_0$ expected to be collected during a time $\Delta T$ and in the whole field 
of view of the Auger observatory with an effective area $S_{eff}$:
\begin{equation}
N(\ge E_0)=S_{eff} \Delta T \int_{0}^{2\pi} d\alpha \int_{\delta_{min}}^{\delta_{max}} d\delta
\int_{E_0} dE J_{tot}(E,\alpha,\delta) \cos \delta ;
\label{eq:numb1}
\end{equation}
using equation (\ref{eq:total1}) and the integrated acceptance of the experiment 
$\mathcal{E}=\Delta\Omega_m S_{eff} \Delta T$, we can rewrite equation (\ref{eq:numb1}) as 
\begin{equation}
N(\ge E_0)=\mathcal{E} \left [ \int_{E_0} dE J_h(E) + \frac{4\pi}{\Delta\Omega_m}
\sum_i \omega(\delta_i) \int_{E_0} dE J_d(E,z_i) \right ].
\label{eq:numb2}
\end{equation}

As we will discuss in the next session, equation (\ref{eq:numb2}) will be used to normalize theoretical 
fluxes to the observations. Once determined its correct normalization to experimental data, we can compute 
the total expected flux in the whole field of view of the Auger observatory simply using equation (\ref{eq:total1}),
so that the total flux will be 

$$ J_{UHECR}(E)=\frac{1}{\Delta\Omega_m}\int_{0}^{2\pi} d\alpha \int_{\delta_{min}}^{\delta_{max}} d\delta 
~J_{tot}(E,\alpha,\delta) \cos \delta =  $$
\begin{equation}
= J_h(E) + \frac{4\pi}{\Delta\Omega_m}\sum_i \omega(\delta_i) J_d(E,z_i)~
\label{eq:total2}
\end{equation}
and the behavior of the expected flux $J_{UHECR}$ can then be directly compared with the observed Auger 
spectrum. The second term in the right hand side of equation (\ref{eq:total2}) depends on the particular
distribution of local sources considered through their declinations, distances (red-shift) and intrinsic luminosity.
In the next session we will discuss peculiar features that appear in the spectrum related to the particular
distribution of local sources considered. 

\begin{figure}[t]
\centering
\includegraphics[bb=50 20 554 373, scale=.5]{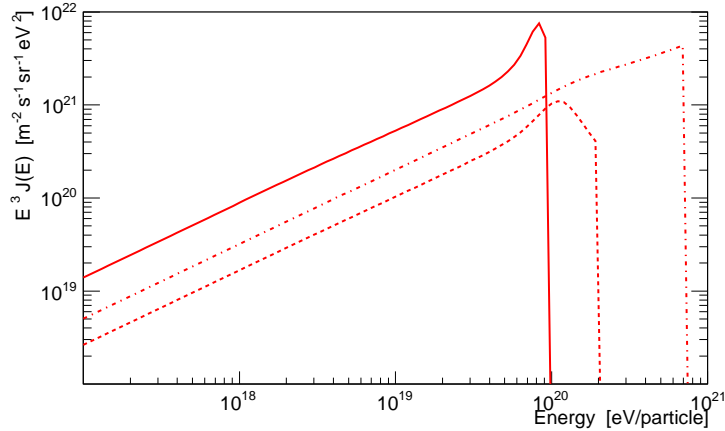} 	
	\caption{\footnotesize{The calculated flux of UHECR (multiplied by $E^3$) from 3 sources of the Auger
	 correlated sources (see text). The curves represent the spectra obtained taking into account the individual
	  characteristics of Cen A ($z=0.001$, $M_B=-14.4$, dot-dashed line), NGC 5506 ($z=0.007$,
	   $M_B=-17.9$, dashed line), NGC 7591 ($z=0.017$, $M_B=-21.6$, full line), the injection parameters 
	   $\gamma=2.2$, $E_{max}=10^{21}$ eV, and cosmological evolution $m=5$.}}    
	\label{fig:example_sources}
\end{figure}

\section{Fluxes}
\label{sec:spectra}

Following the analysis developed by the Auger collaboration, we will take the local sources from the 12th edition
of the VCV catalog which includes the sources that contribute to the observed anisotropy in the Auger events.
The correlation signal claimed by the Auger collaboration is based on the fact that the directions of 20 out of 27
UHECR events \cite{AugerCorrApp} collected between January 1 2004 and August 31 2007 with $E \geq 57$
EeV are correlated (within $\psi_0=3.2^{\circ}$) with the directions of astrophysical objects 
(with $z \leq z_0 = 0.017$)
in the VCV catalog\footnote{In this paper we use the first Auger collaboration result on correlation since the list of
correlated events reported in \cite{AugerCorrApp} is the only one available in literature.}. 
It means that corresponding to the direction of 20 UHECR events there is at least one source in the catalog with
these selection parameters. To select the actual source we consider the source power, choosing the most
luminous (taking the absolute magnitude in the B band) corresponding to the direction of an individual UHECR
event. Therefore we are implicitly assuming that UHECR luminosity is proportional to the photon luminosity. 

In figure (\ref{fig:example_sources}) we show the calculated fluxes of three sources of this list as an example. 
One can see the difference between the cut-off due to the different red-shifts of the AGNs, the fluxes being
calculated with $E_{max}=10^{21}$ eV and with the intrinsic luminosity of the source from the catalog. Our
sample includes sources with absolute magnitude (in the B band) in the range $-21.6<M_B<-14.4$, that
corresponds to photon luminosity in the range $4.5\times10^{40}<L<3.4\times10^{43}$ erg/s, with a mean value
of $8.5\times 10^{42}$ erg/s.

\begin{figure}
\centering
\includegraphics[bb=50 20 554 373, scale=.5]{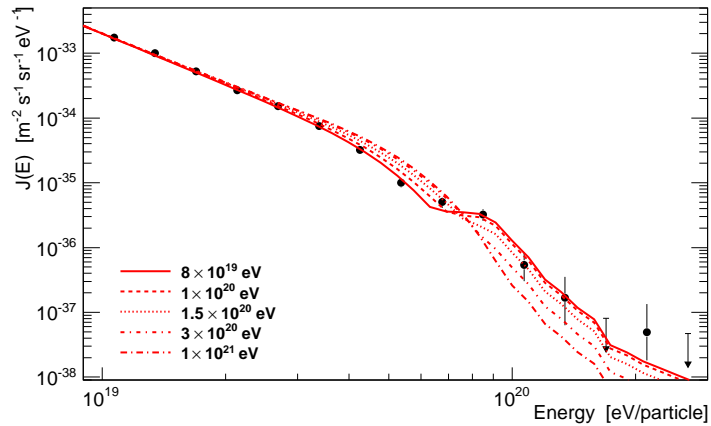} 	
	\caption{\footnotesize{The calculated flux of UHECR as the sum of the two contributing populations
	with different values of the maximum injection energy for the homogeneous distribution, compared with 
	the Auger spectrum (the upper limits correspond to 68$\%$ CL).
	The local sources are the VCV objects correlated with Auger highest energy events (see text).
	The maximum energy for the local sources is fixed at $10^{21}$ eV. All the models have been normalized
	fixing the observed number of events as specified in the text.}}
	\label{fig:Emax}
\end{figure}

Among injection parameters, the  evolution constant $m$ and the injection spectral index $\gamma_g$ affect 
the shape of the spectrum to a large extent. A fit procedure to determine their best values, separating the two
components, is difficult and rather uncertain because of their correlation. For simplicity we then assume the pair
($\gamma_g$, $m$) to be the same for the two source populations. It has to be pointed out that under this
assumption the choice of the parameters is strongly guided by low energies ($E < 20 \div 30$ EeV) and 
therefore the parameter values are mainly fixed comparing theoretical spectra and Auger data in this energy region.

In the present paper we compare our theoretical results with the Auger spectrum released in 2009
\cite{icrc09Schu,SpectrumPLB}. To allow a better agreement with the dip model \cite{dip} the Auger energy 
scale has been shifted towards higher energies by 20\%
\footnote{
Hereafter we will use for the Auger data this new energy scale.}. 
This shift is compatible with the systematic uncertainties in the energy determination quoted by the Auger
collaboration and is also supported by a recent analysis of the muon content in the Auger data \cite{icrc09Cas},
which indicate an energy scale increased by about this factor. Using the Auger spectrum with the shifted energy
scale we find a good agreement with our model in the scenario with a strong cosmological evolution ($m=5$)
and an injection power law index $\gamma_g=2.2$. In the following we will always keep fixed these two 
parameters at the values quoted here, this assumption roughly agrees also with the recent spectrum analysis
performed by the Auger collaboration \cite{icrc09Schu}. 

To normalize the theoretical flux we determine the number of events corresponding to the
Auger exposure equating to experimental data. Since we consider two populations of sources, we have to
integrate separately the two fluxes in the high and low energy ranges. In the lower energy range 
($3.8 < E < 48$ EeV) the total flux is dominated by the homogeneous distribution, while in the higher 
energy range ($60 < E < 240$ EeV) the local sources could play the leading role. 

The crucial parameters that drive the flux behavior at the transition from a homogeneous to a local distribution
of sources are the maximum energy $E_{max}$ and the turning redshift $z_0$. In the present paper we have 
considered a twofold analysis: fixing the value of $z_0$, as in the Auger analysis, and varying the maximum energy 
of acceleration $E_{max}$ or fixing the maximum energy varying the turning redshift $z_0$. In the two following 
subsections we will discuss separately the results obtained under the two assumptions.

\subsection{Maximum Energy Analysis} 
Let us start with the case in which the value of the turning redshift between local and homogeneous sources 
$z_0$ is fixed as in the Auger analysis. In this case $z_0=0.017$ corresponds to the redshift of the farther 
correlated source. This analysis assumes that the local discrete sources are only the sources that in the VCV 
catalog show a correlation with the Auger events. 
The best value for $E_{max}$ of the local sources is $10^{21}$
eV. In figure (\ref{fig:Emax})  we show a comparison of the Auger data with theoretical spectra obtained through
different choices of $E_{max}$ for the homogeneous
distribution of sources; the fluxes are normalized fixing the observed number of events as discussed above. 
The comparison with data above $3 \times 10^{18}$ eV gives a $\chi^2/ndf$ of 0.45, 0.64, 1.3, 2.3, 2.7 for 
$E_{max} = 8 \times 10^{19}, 1 \times 10^{20}, 1.5 \times 10^{20},  3 \times 10^{20}$ and 
$ 1 \times 10^{21}$ eV respectively.

\begin{figure}
\centering
\includegraphics[bb=50 20 554 373, scale=.5]{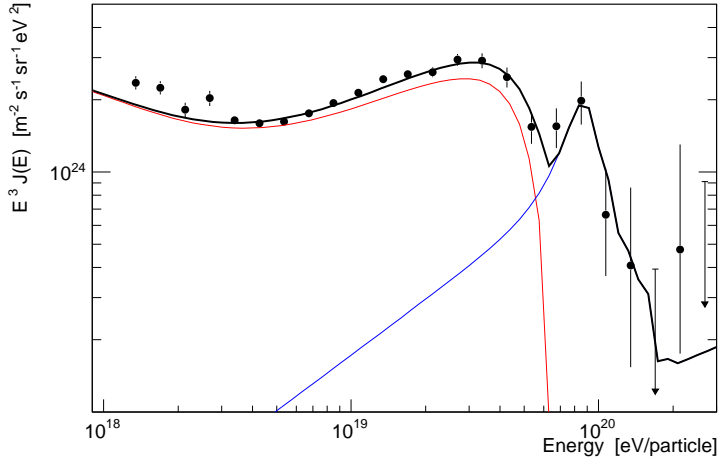} 	
	\caption{\footnotesize{The calculated flux of UHECR as the sum (black line) of the two contributing
	populations multiplied by $E^3$. The local sources are the VCV objects correlated with Auger (see text).
	 $E_{max}$ is $8\times 10^{19}$ eV for the homogeneous distribution (red line),  $10^{21}$ eV for the local
	  sources (blue line).}}
	\label{fig:ourbestspectrum}
\end{figure}

In order to develop our analysis we choose therefore the value of $E_{max}$ for the homogeneous distribution
that shows the best agreement with data: $E_{max_{h}}=8\times10^{19}$ eV. 
It has to be noted that this choice is strongly driven by the small excess in the Auger spectrum data at 
$8.4 \times10^{19}$ eV ($7 \times10^{19}$ eV with original energy scale). It lies at about $(1.5 \div 2) \sigma$
above the functions used to fit the spectrum \cite{SpectrumPLB}, so the statistical significance of the excess is
still rather poor and only new data can confirm or deny it.

We use as local sources the ones associated to the Auger correlated events. This list, obtained with the
luminosity criterion described above, includes 17 independent sources since three of them are associated to two
events. We use this set of independent sources. However we observe that even taking the 20 sources (with 
three repetitions among them) there are no substantial changes in the calculated
flux apart minor differences at the highest energies, where statistics is still too poor to discriminate among 
different choices. This evidence is a direct consequence of the normalization procedure which recovers the 
lack of sources readjusting their luminosity. The spectrum obtained with the 17 independent 
sources of the VCV catalog is shown in figure (\ref{fig:ourbestspectrum}).
 
As discussed in the last section, the normalization procedure fixes the emissivity of the homogenous 
distribution of sources $(z>z_0)$ and the luminosity of the local distributed point sources $(z<z_0)$.
In particular, in the case of local sources we have assumed that the UHECR luminosity is proportional 
to the photon luminosity (taken from the catalog) with a single conversion factor common to all sources
(our actual normalization parameter). Taking into account the number of local contributing
sources and their distance from us we can easily determine the number density associated to these sources. 
Using the average UHECR luminosity of the local sources and their number density we can 
determine a reference value for the emissivity of local sources comparing it with the emissivity of the
homogenous distribution, which is directly fitted from data. We obtain an emissivity of 
$1.9\times 10^{39}$ erg/(s$\mathrm{Mpc}^3$) for the homogeneously distributed sources and a value of
$2.4\times 10^{39}$ erg/(s$\mathrm{Mpc}^3$) for the reference emissivity of local sources. 

\begin{figure}
\begin{minipage}[t]{8cm}
\centering
\includegraphics[bb=50 20 554 373, scale=.4]{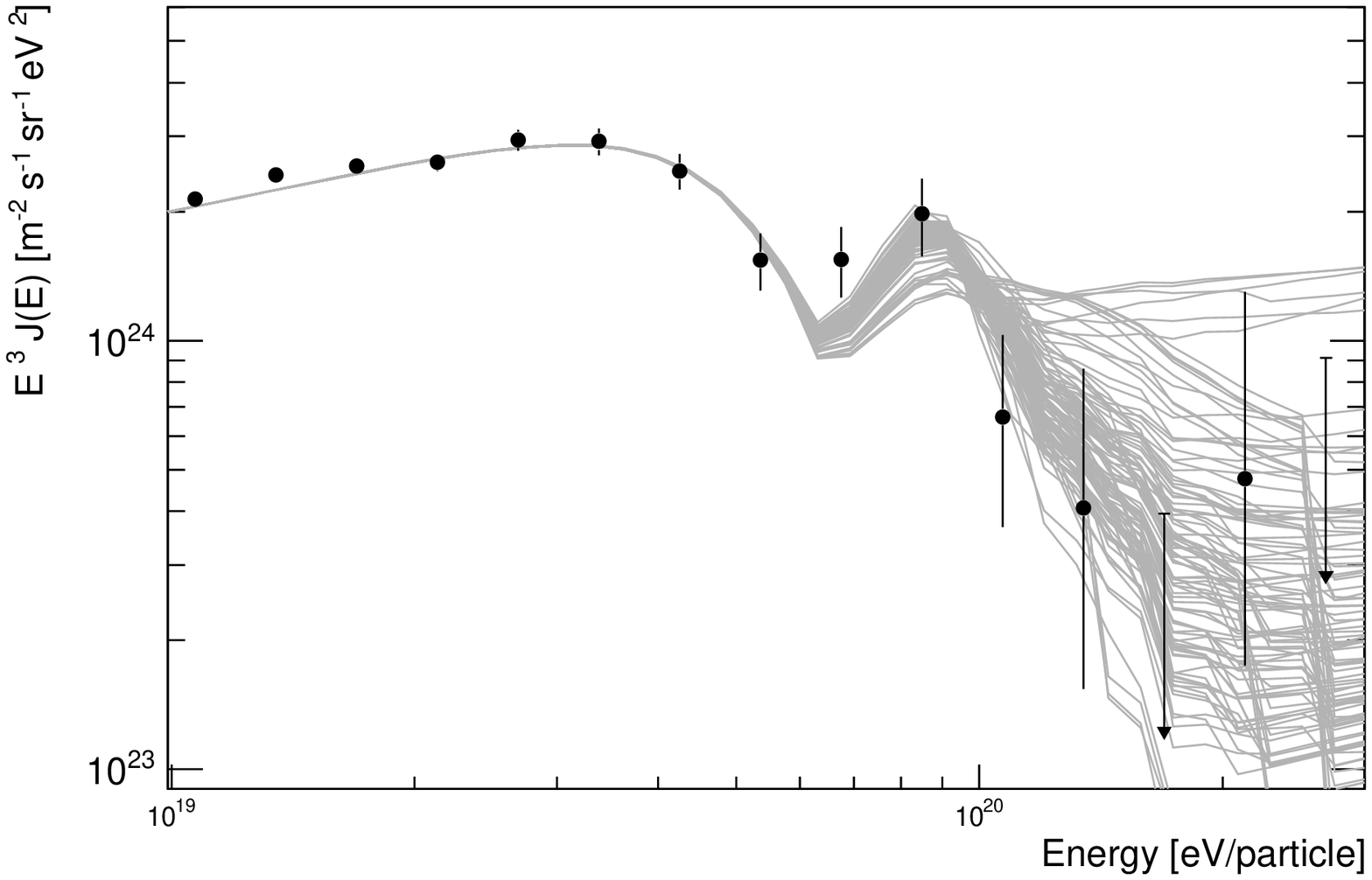}
\end{minipage}
\ \hspace{1mm} \hspace{1mm} \	
\begin{minipage}[t]{8cm}
\includegraphics[bb=50 20 554 373, scale=.4]{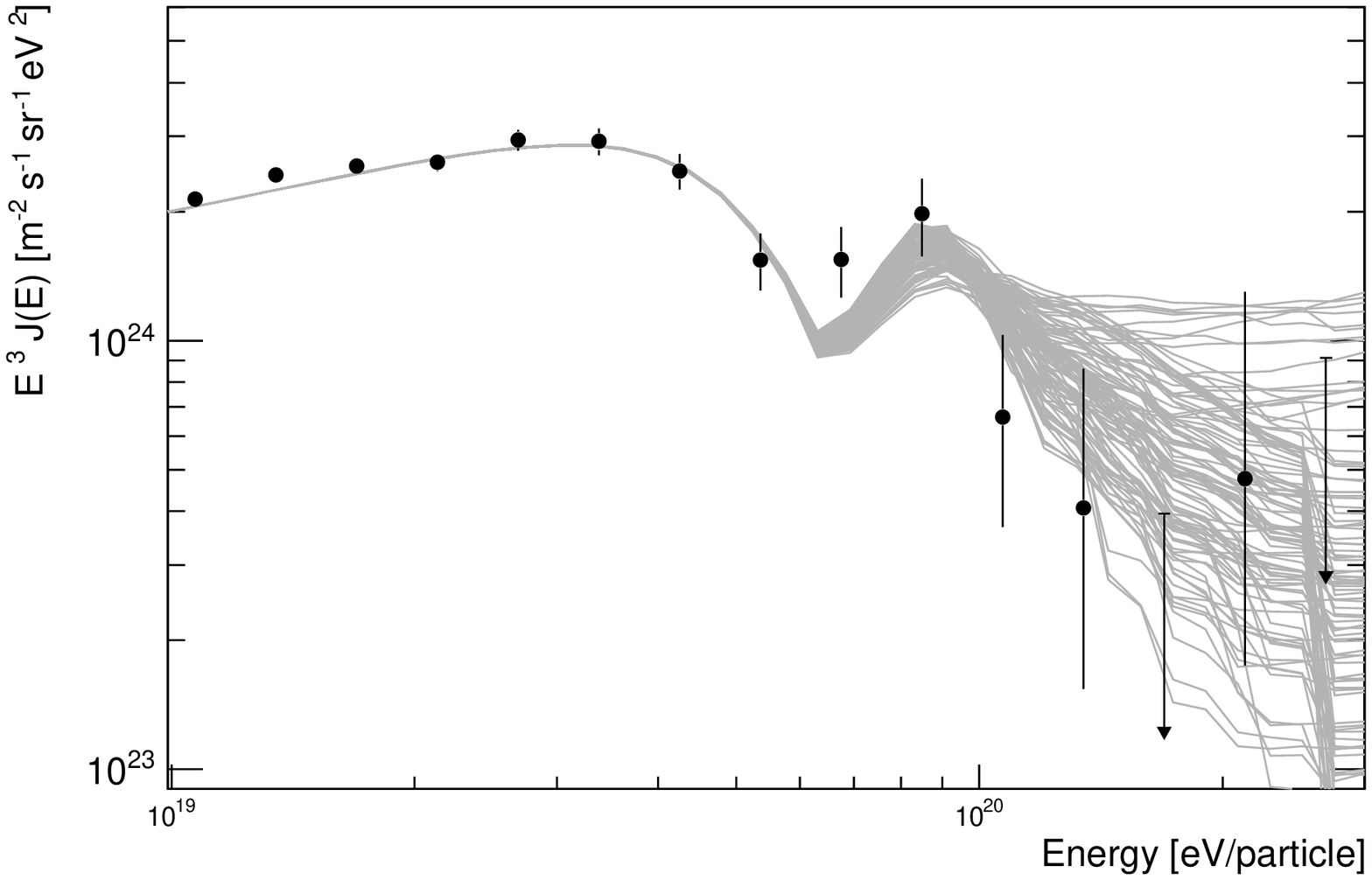}
\end{minipage}
	\caption{\footnotesize{Left panel: the spectrum of UHECR multiplied by $E^3$ of 100 lists made of the 17
	 correlated sources and 8 sources extracted by the VCV catalog (with red-shift $\leq0.017$ and in the field
	  of view of the Auger observatory) and with directions extracted from an isotropic distribution of sources in 
	  the field of view of the Auger observatory. Right panel: the spectrum of UHECR multiplied by $E^3$ of 
	  100 lists made of the 17 correlated 
	  sources and 25 sources, with the same procedure of the left panel.}}    
	\label{fig:100lists_cor}
\end{figure}

The published Auger result on correlation \cite{AugerCorrSc,AugerCorrApp} refers to an exposure 
sensibly lower\footnote{
It has to be remarked that the difference in the exposures is not only due to the period of data taking, but 
also to the different selections cuts.} 
(by about 30\%) than in the recent published spectrum \cite{icrc09Schu,SpectrumPLB}, used here as reference.
In the latter data, the number of collected events above 60 EeV (50 EeV with original energy scale) is 59 against
the 17 sources we use in our calculation. The analysis of event multiplets is completely outside the aims of this
paper, because we are only interested in the spectrum features. Neverthless one should be aware of the fact that
the 17 correlated sources cannot account alone for the observed high energy events. In ref. \cite{AugerCorrApp}
the Auger collaboration sets a lower limit on the number of sources obtained from the
ratio of doublets to singlets and based on the simple assumption that all the sources have the same
apparent luminosity \cite{Dubovsky}; this limit turns out to be 61 sources. Then using our set of sources we
are implicitly assuming that each source contributes with more than one event to the flux. 
The assumption we made to agree with the lack of multiplets is that the missing sources have
the same astrophysical properties as the correlated sources and so they are representative of the whole set of
``true'' sources.   

Let us now check this assumption and investigate how the characteristics of the missing sources can influence
the calculated flux. We consider here a set of sources made of the 17 correlated 
sources and of $N_{add}$ added sources, chosen from the VCV catalog (with their characteristic values of 
red-shift and luminosity) in the field of view of the Auger observatory and with $z\leq 0.017$. The directions of the 
$N_{add}$ sources have been then replaced with directions extracted from an isotropic distribution in the 
field of view of the Auger observatory.
The choice of $N_{add}$ is fixed by the correlation signal: the events from the 17 local sources are correlated by
construction.  $N_{add}$ is then set to 8 or 25 to agree with the published correlation results of $70\%$ and 
$40\%$ in \cite{AugerCorrSc} and \cite{Auger_corr_up} respectively.

The results are shown in figure (\ref{fig:100lists_cor}).
From the comparison we observe that substantial changes in the calculated flux are visible only at the highest
energies, where we can not discriminate among different lists because of the data statistics; we also observe 
that there are not changes between the example with $N_{add}=8$ and $N_{add}=25$ because the
normalization procedure readjusts the flux.  

The analysis presented here assumes that the maximum energy of acceleration changes with the 
class of sources, with a larger maximum energy associated to the local sparse sources and a lower one 
for the homogeneously distributed. While this assumption enables a very good fit of the experimental data 
seems poorly justified from an astrophysical point of view.

\begin{figure}
\begin{minipage}[t]{8cm}
\centering
\includegraphics[bb=50 20 554 373, scale=.4]{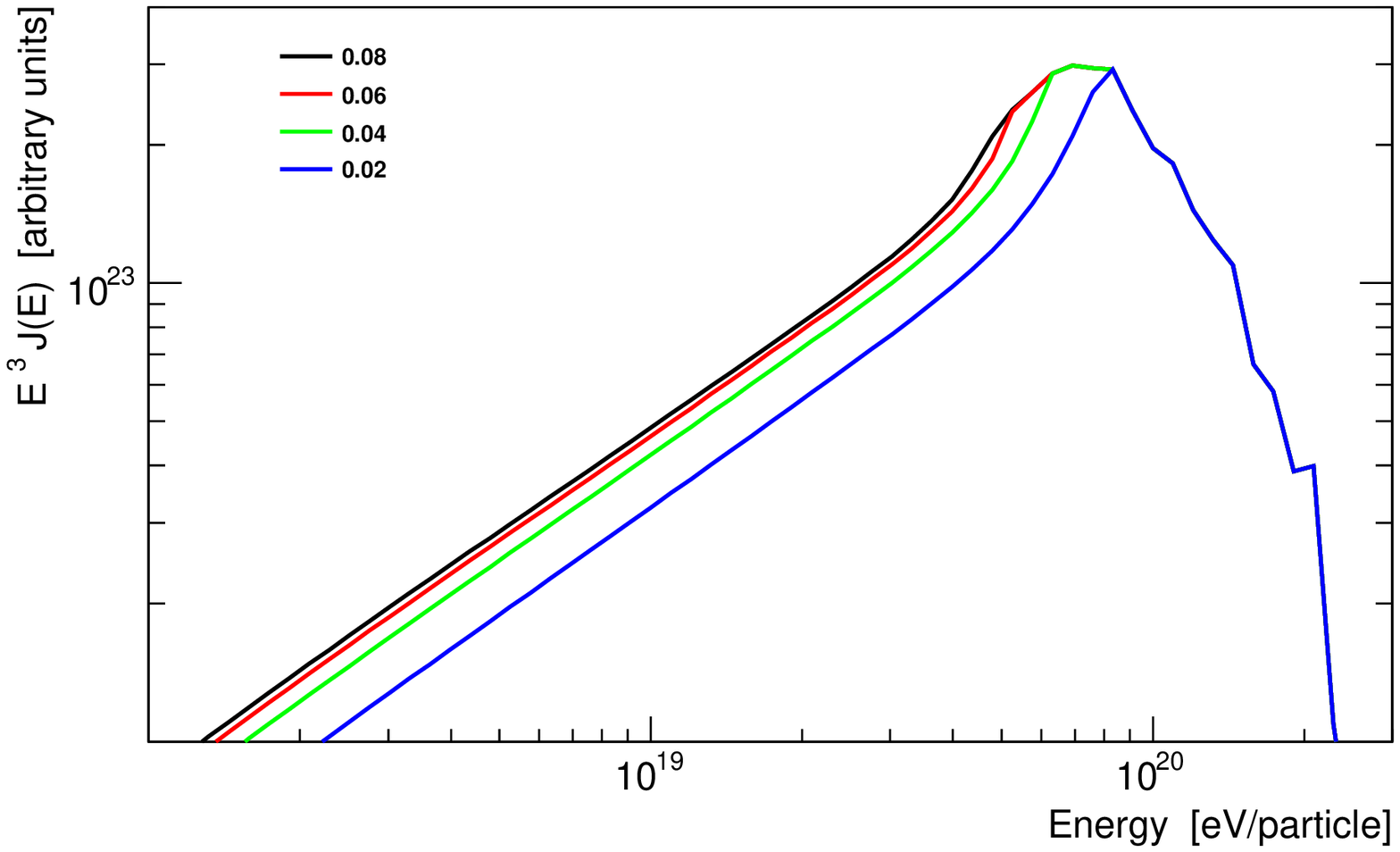}
\end{minipage}
\ \hspace{1mm} \hspace{1mm} \	
\begin{minipage}[t]{8cm}
\includegraphics[bb=50 20 554 373, scale=.4]{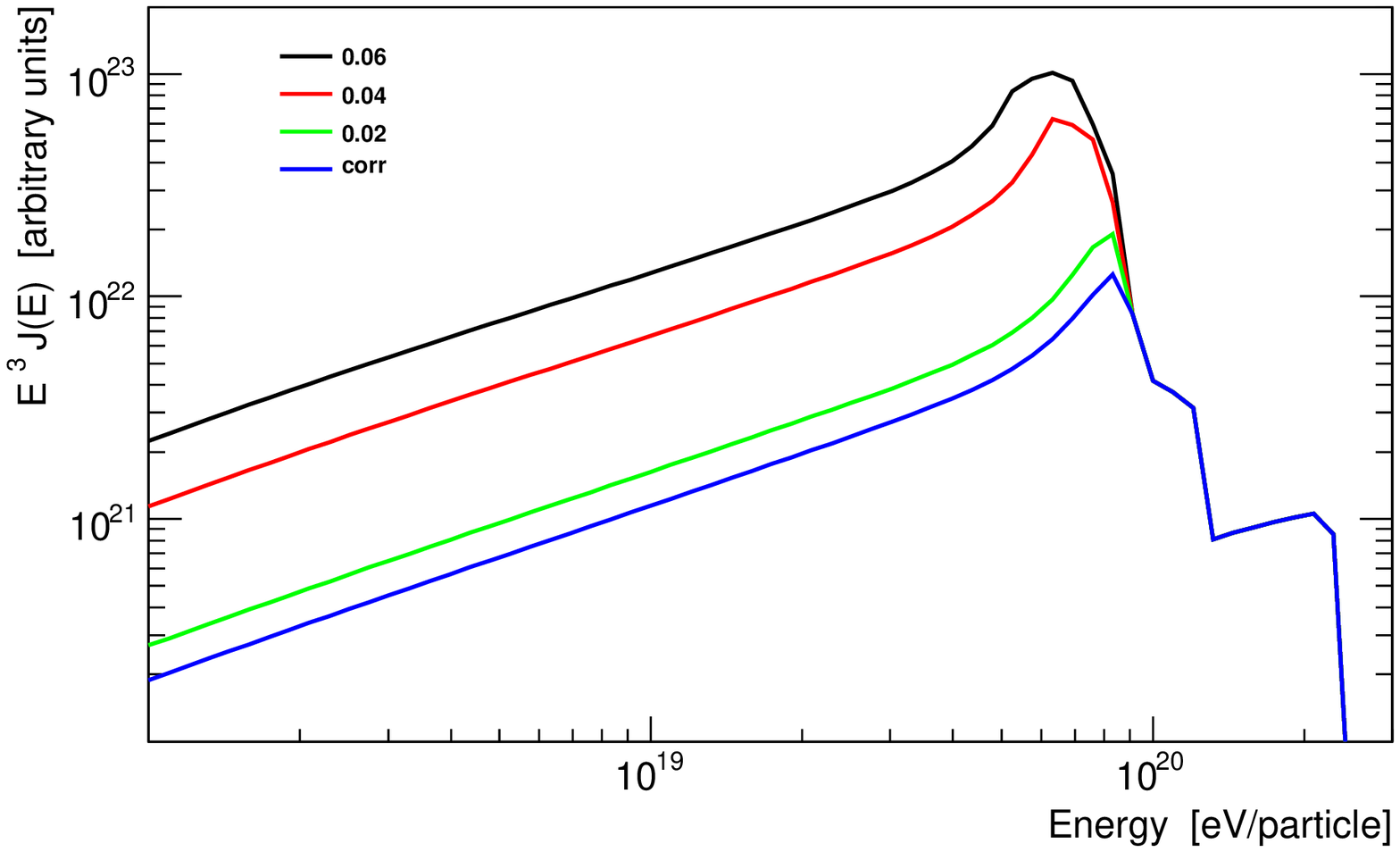}
\end{minipage}
	\caption{\footnotesize{Left panel: UHECR flux from the complete set of sources which fall in the Auger field
	 of view taken from the VCV catalog at redshift less than $z_0=0.02,0.04,0.06, 0.08$ (as labelled). Right panel:
	  UHECR flux from one set of VCV sources made of the 17 correlated sources and $N_{add}$ sources extracted
	   from the VCV catalog in the field of view of Auger (with redshift $z<z_0$ as labelled); $N_{add}$ is chosen to
	    keep the density $n_{corr}$ fixed with redshift. The curve labelled $corr$ corresponds to the flux of the 
	Auger correlated sources with $z_0=0.017$.}}
	 \label{fig:manyz0}
\end{figure}

\subsection{Turning Redshift Analysis} 
 The analysis presented so far takes the parameter $z_0$ fixed as in the Auger study of correlation 
($z_0=0.017$); in this case, as follows from the discussion above, the only relevant parameter to fit the 
spectrum is the maximum energy $E_{max}$ associated to local sparse sources and uniform distributed 
sources. Let us now change our point of view using $z_0$ as a free parameter to fit the observed spectrum; 
in this case we will keep the maximum energy fixed to the same value for both local and faraway sources. As
discussed in the previous section the possible value of $z_0$ could not be too different from the so-called 
GZK-sphere, the distance inside which UHE protons are less affected by the photo-pion production process. 
This typical distance depends on the particles energy: for a proton of $E>50$ EeV it lies in the range of 
$100\div 300$ Mpc. Therefore the values of $z_0$ we are interested in lay in the range $z_0\simeq 0.01\div 0.1$. 

Changing the parameter $z_0$ respect to the Auger analysis of correlation introduces a new uncertainty in 
our analysis, because we are not anymore guided by the Auger result in choosing the contributing local sources. 
The local AGN as listed in the VCV catalog are much more numerous respect to the list of $17$ sources whose
position shows a correlation with the Auger events. Moreover, as pointed out above, the possible values of $z_0$ 
are relatively low with $z_0\le 0.1$ which corresponds to a distance of around $400$ Mpc. Inside this radius the
VCV catalog can be considered quite accurate being a complete sample of the local Universe. 

\begin{figure}
\begin{minipage}[t]{8cm}
\centering
\includegraphics[bb=0 0 567 381, scale=.4]{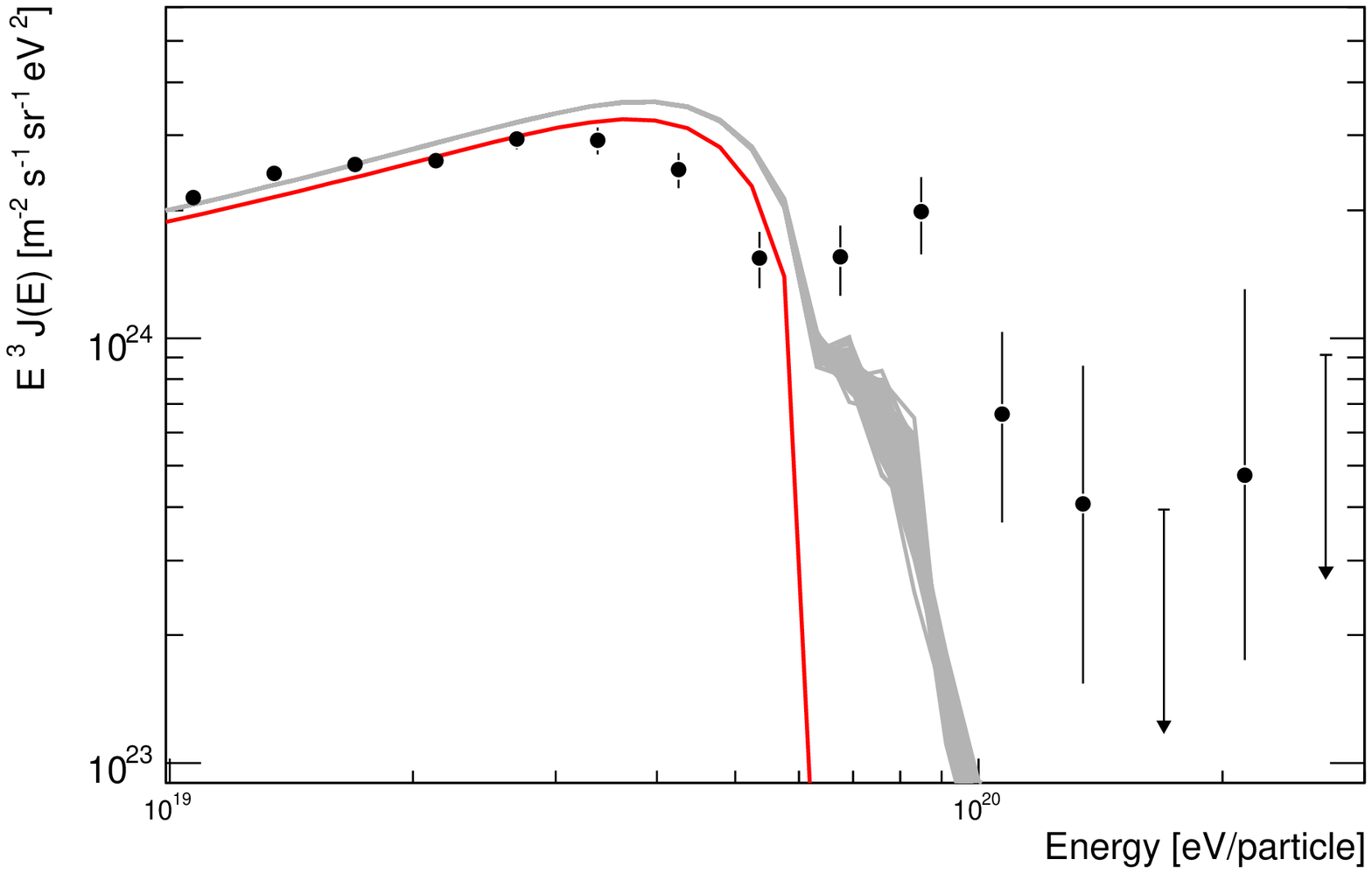}
\end{minipage}
\ \hspace{1mm} \hspace{1mm} \	
\begin{minipage}[t]{8cm}
\includegraphics[bb=0 0 567 381, scale=.4]{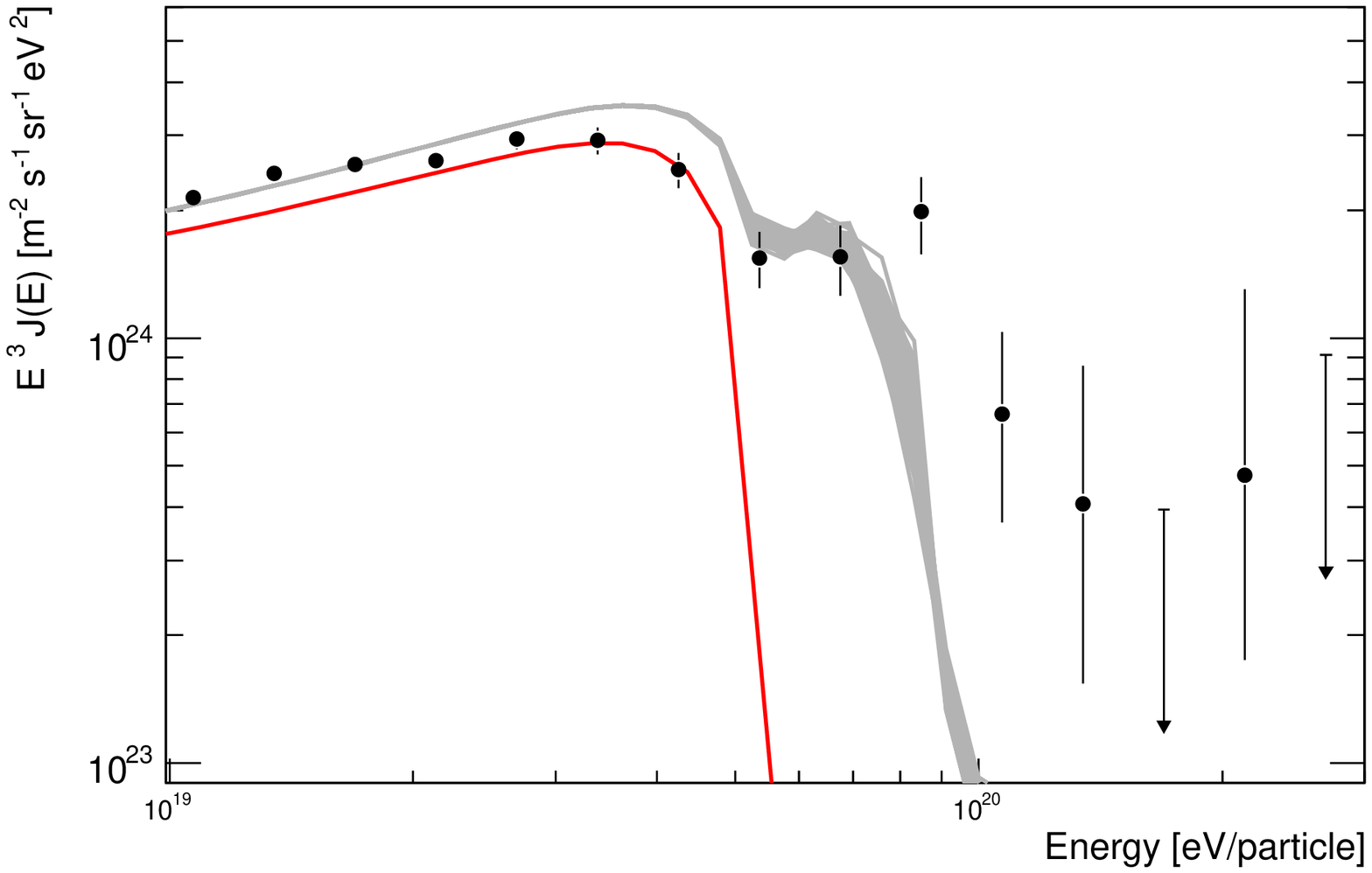}
\end{minipage}
	\caption{\footnotesize{The spectrum of UHECR (multiplied by $E^{3}$) produced by an homogeneous
	 distribution of sources at $z>z_0$ (red continuos line) and local discrete sources at $z<z_0$. The local 
	 sources correspond to 100 lists made of the $17$ Auger correlated sources at $z<0.017$ and $N_{add}$
	  sources in the field of view of Auger at $0.017\le z\le z_0$ extracted from the VCV catalog with fixed 
	  density $n_{corr}$ (see text). Left panel $z_0=0.04$ and right panel $z_0=0.06$.}}
	\label{fig:flux_n_corr}
\end{figure}

To solve the ambiguity in the choice of local contributing sources we have considered two different recipes. 
In the first recipe we have fixed the local density of sources assuming that it corresponds to the density of the $17$
Auger correlated sources: $n_{corr}=10^{-5}$ Mpc$^{-3}$. In this case in the redshift range $0\le z \le 0.017$ we 
have used only the $17$ Auger correlated sources and at larger redshift $0.017\le z \le z_0$ we have added a set 
of $N_{add}$ sources, chosen from the VCV catalog in the field of view of the Auger observatory, with the 
same density $n_{corr}$ of the local correlated sources. In the second recipe we have simply taken the all 
VCV sources inside $z_0$, without assigning any special role to the sources tagged in the Auger analysis. 

In order to understand the effect of the parameter $z_0$ on the flux from local sources, in figure
\ref{fig:manyz0} we show the flux computed using the sources of the VCV catalog placed at red-shift $z<z_0$ 
for different values of the parameter $z_0$ (as labelled) and using the two recipes introduced above. In the 
right panel of figure \ref{fig:manyz0} we have fixed the density of the local sources at $n_{corr}$ (first recipe);
the lowest curve corresponds to the flux from the $17$ Auger correlated sources within $z_0=0.017$ used 
in the previous analysis. In the left panel of figure \ref{fig:manyz0} we have used the all VCV 
sources inside $z_0$ as labelled (second recipe). From figure \ref{fig:manyz0}, independently of the assumptions on 
the local density of sources, it is evident that increasing the number of the discrete
sources, by taking larger values of $z_0$, the expected contribution of the local (discrete) universe becomes more
relevant at the lowest energies with a gradual shift of the flux maximum toward lower energies.

The flux of the homogeneous distributed sources also depends on the choice of $z_0$. In our analysis we have 
assumed the homogenous distribution characterized by a cut-off in redshift with a null injection at $z<z_0$
(see equation (\ref{eq:inj_h})). Therefore, increasing $z_0$ it is reduced the size of the Universe contributing 
with homogeneous sources and the corresponding contribution to the total flux will be reduced. This reduction 
is expected at the highest energies where the particles contributing to the flux are produced nearby. In other 
words the flux of homogeneous sources will show a suppression at high energy with a reduced value of the 
cut-off energy increasing $z_0$.

\begin{figure}
\begin{minipage}[t]{8cm}
\centering
\includegraphics[bb=50 20 554 373, scale=.4]{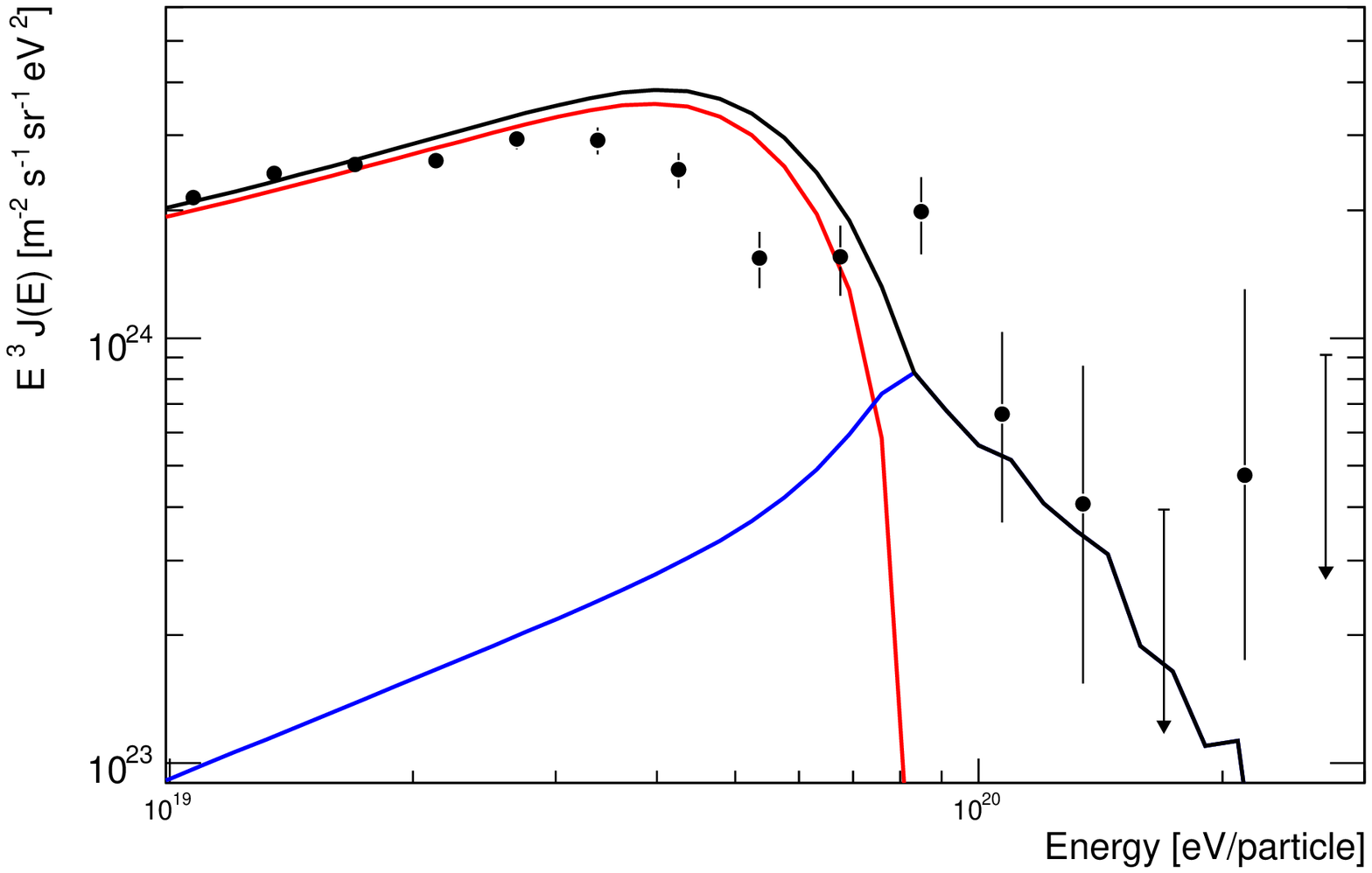}
\end{minipage}
\ \hspace{1mm} \hspace{1mm} \	
\begin{minipage}[t]{8cm}
\includegraphics[bb=50 20 554 373, scale=.4]{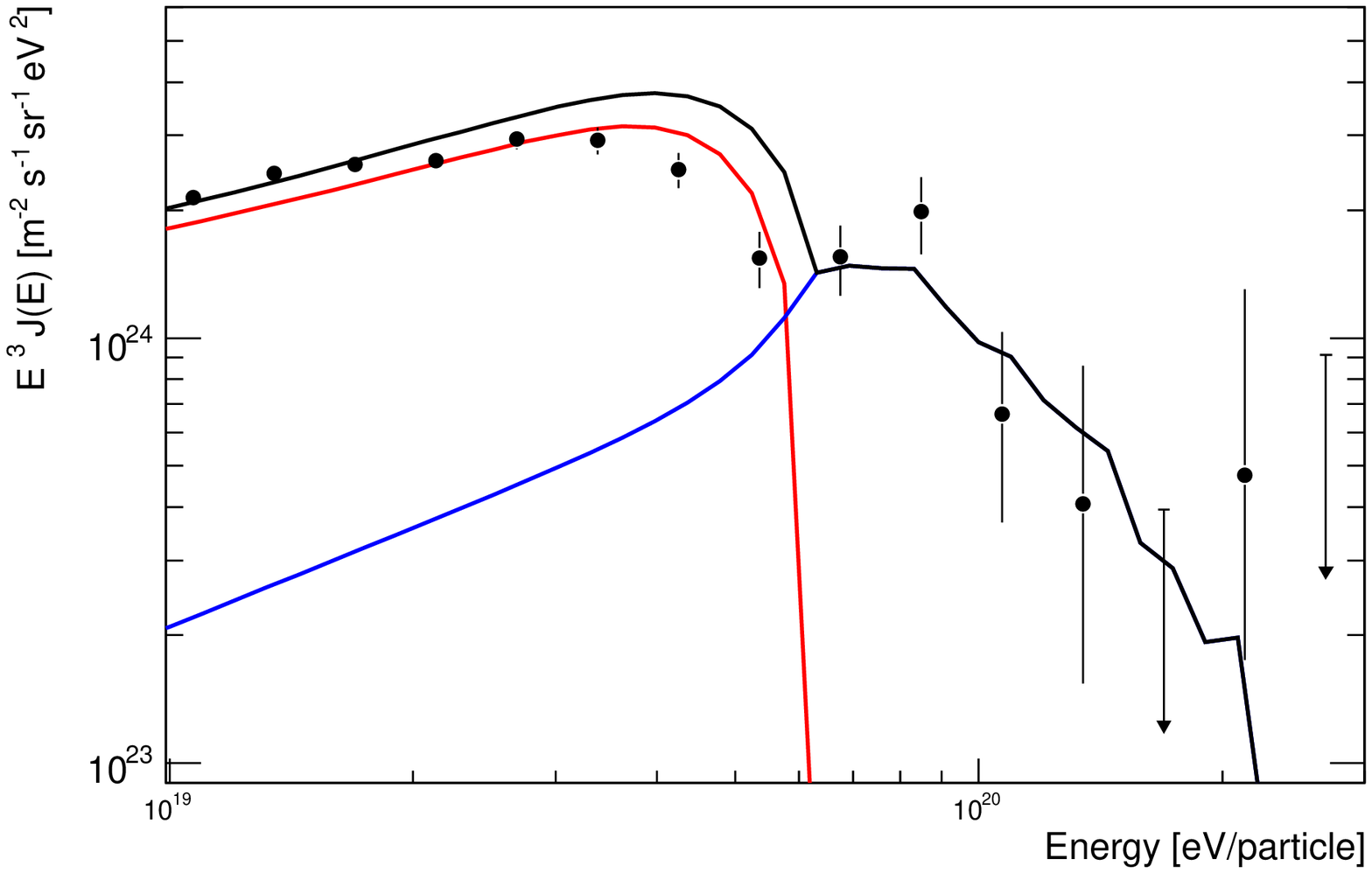}
\end{minipage}
	\caption{\footnotesize{The spectrum of UHECR (multiplied by $E^{3}$) produced by an homogeneous
	 distribution of sources at $z>z_0$ (red continuos line) and local sources at $z<z_0$. The local 
	 sources correspond to all VCV sources inside $z_0$ and in the field of view of Auger. Left panel $z_0=0.02$
	 and right panel $z_0=0.04$.}}
	 \label{fig:flux_z0}
\end{figure}

In the following analysis we will consider the same source model for all sources: homogeneous and discrete, 
fixing, as before, an injection power law $\gamma_g=2.2$, an evolution parameter $m=5$ and assuming the 
same maximum energy $E_{max}=3\times 10^{20}$ eV for all sources. 

In figure \ref{fig:flux_n_corr} we show the flux computed using, as discussed above, our first recipe: fixing the 
contributing sources at $z\le 0.017$ to the $17$ Auger correlated sources and assuming that the density 
$n_{corr}=10^{-5}$ Mpc$^{-3}$ remains constant with redshift. In figure \ref{fig:flux_n_corr} the value
of $z_0$ is fixed to $z_0=0.06$ (right panel) and to $z_0=0.04$ (left panel). The red continuos line shows the
contribution of the homogenous sources. The contribution of the local sources, shadowed band, is characterized by
the uncertainty in the choice of the sources in the redshift range $0.017\le z \le z_0$ with 100 different possible
lists of VCV sources in the field of view of Auger contributing to the flux. The best fit
emissivities for homogeneous and local sources are respectively $2\times 10^{39}$ erg/(sMpc$^3$) and
$5.6\times 10^{38}$ erg/(sMpc$^3$) for $z_0=0.06$, and $5\times 10^{38}$ erg/(sMpc$^3$) in the case of
$z_0=0.04$. 

As follows from figure \ref{fig:flux_n_corr} the highest energy part of the observed flux cannot be well 
described by the theoretical spectrum. The disagreement at the highest energies can be understood
taking into account the effects of the sources added at redshift larger than $0.017$. These sources increase 
the contribution to the flux of the local universe at low energy and, consequently, increase the jump in the flux
between its maximum and minimum values. These effects, together with the relatively low maximum energy
$E_{max}=3\times 10^{20}$ eV, imply that the highest energy part of the observed flux cannot be well 
described by the theoretical spectrum. 

A better agreement with Auger data can be obtained following our second recipe in which the 
information on the $17$ Auger correlated sources is neglected. In this case as discrete local sources 
we have considered the complete set of sources in the field of view of Auger present in the VCV catalog at $z<z_0$.
In figure \ref{fig:flux_z0} we plot the flux obtained with two different values of $z_0=0.02, 0.04$, keeping the injection
parameters fixed as above ($E_{max}=3\times 10^{20}$ eV, $\gamma_g=2.2$ and $m=5$) for all sources. 
The number of sources of the VCV catalog in the field of view of Auger inside $z_0=0.02$ is $346$, with an
associated density of $1.3\times 10^{-4}$ Mpc$^{-3}$; in the case of $z_0=0.04$ the number of VCV sources is
$835$ with a density $4\times 10^{-5}$ Mpc$^{-3}$. As follows form figure \ref{fig:flux_z0} the agreement
with experimental data at the highest energies is improved. The source emissivities used to fit the Auger data 
in figure \ref{fig:flux_z0} are $2\times 10^{39}$ erg/(s Mpc$^3$) for the homogeneous sources and 
$6.5\times 10^{38}$ erg/(s Mpc$^3$), $4.5\times 10^{38}$ erg/(s Mpc$^3$) respectively in the case of $z_0=0.02$ 
and $z_0=0.04$.

The result of figure \ref{fig:flux_z0} is the effect of many sources present in the VCV catalog that stands nearby the
observer. In the field of view of Auger there are around $60$ sources of the VCV catalog inside $z<0.005$. As follows
from the Auger analysis these sources do not show any correlation signal nevertheless if added at the contributing
sources of UHECR could provide a better description of the observations at the highest energies. 

\begin{figure}
\begin{minipage}[t]{8cm}
\centering
\includegraphics[bb=50 20 554 373, scale=.4]{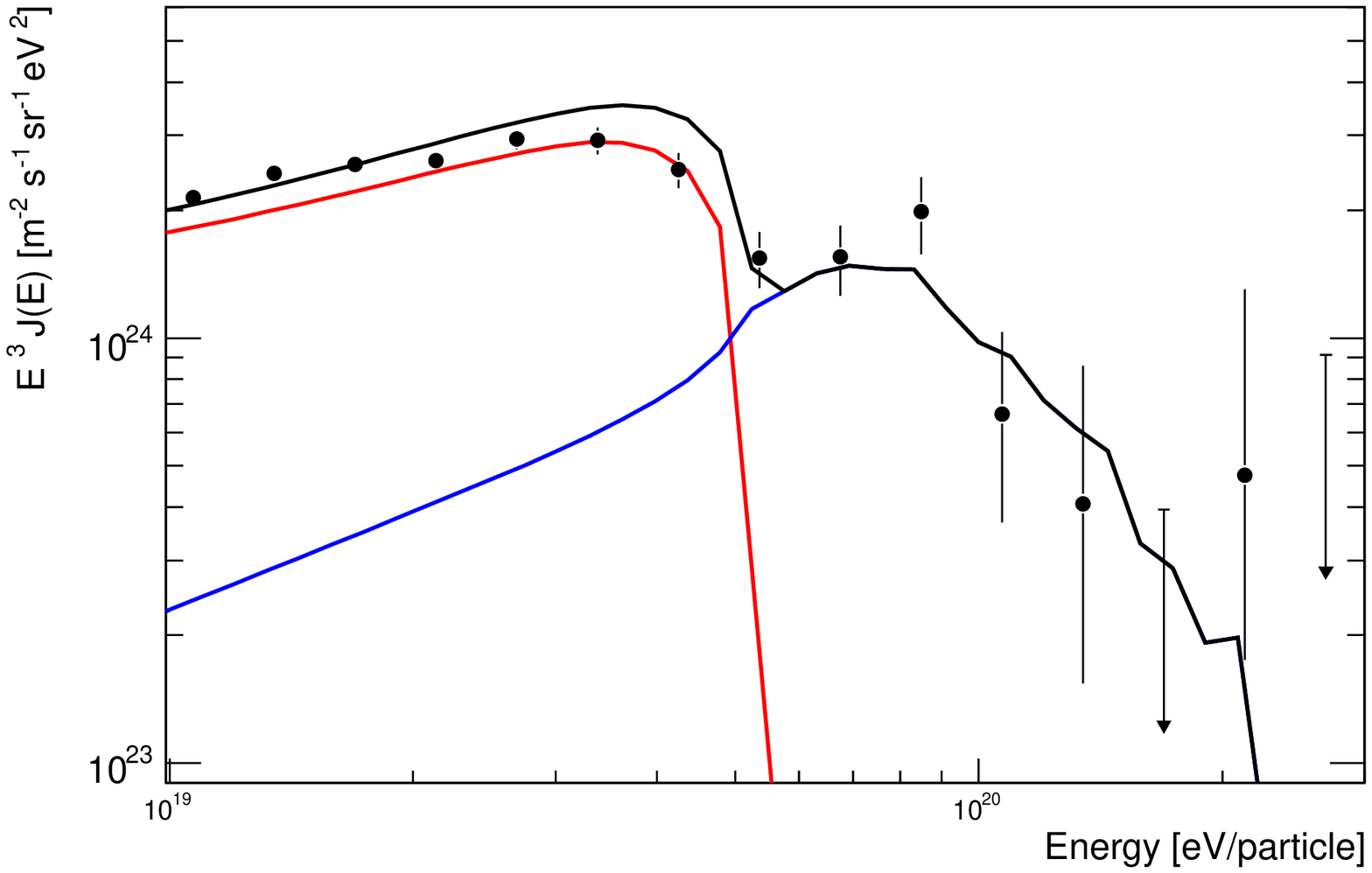}
\end{minipage}
\ \hspace{1mm} \hspace{1mm} \	
\begin{minipage}[t]{8cm}
\includegraphics[bb=50 20 554 373, scale=.4]{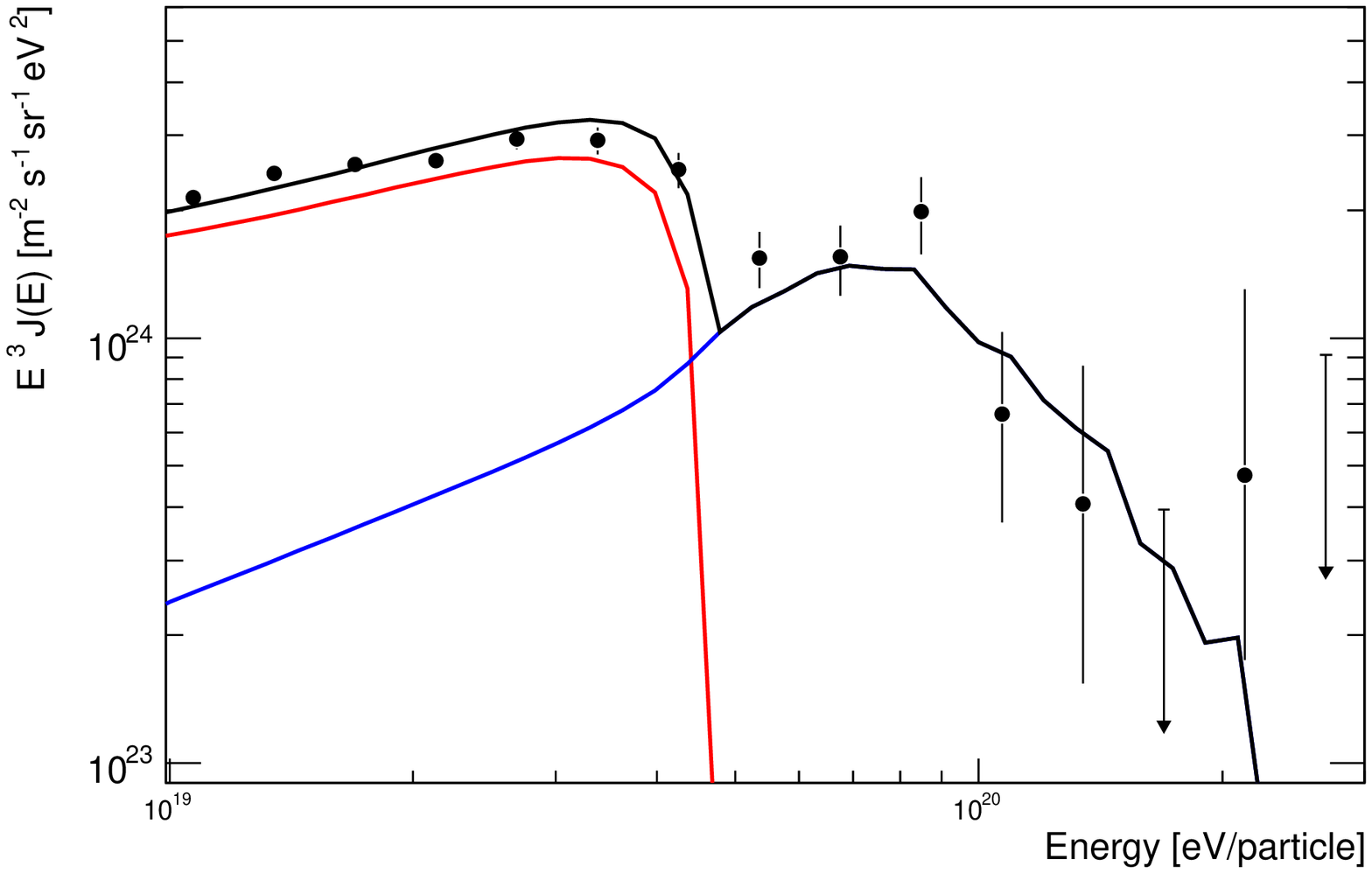}
\end{minipage}
	\caption{\footnotesize{The same as in figure \ref{fig:flux_z0} with $z_0=0.06$ (left panel) 
	and $z_0=0.08$ (right panel).}}
	 \label{fig:flux_z0_006}
\end{figure}

In figure \ref{fig:flux_z0_006} we have repeated the same analysis of figure \ref{fig:flux_z0} fixing the parameter
$z_0=0.06$ (left panel) and $z_0=0.08$ (right panel). In the case of $z_0=0.06$ the number of sources of the
VCV catalog is $1379$ with a density of $2\times 10^{-5}$ Mpc$^{-3}$, while for $z_0=0.08$ the number of involved
sources is $2064$ with a density of $1.3\times 10^{-5}$ Mpc$^{-3}$. The emissivities of the uniform and discrete
components, fitted from Auger data, are $2\times 10^{39}$ erg/(s Mpc$^3$) for homogeneous sources and 
$3\times 10^{38}$ erg/(s Mpc$^3$) for discrete sources with $z_0=0.06$ and $1.6\times 10^{38}$ erg/(s Mpc$^3$) 
in the case of $z_0=0.08$. The agreement among theoretical and observed spectra with $z_0=0.06, 0.08$ is further
improved with the contribution of the homogeneous distributed sources restricted to lower energies respect to 
the case of lower $z_0$. In the case of $z_0$ values larger than $0.08$ the agreement with experimental 
data starts to become less good because of the substantial cut of the flux from homogeneous sources at the 
lowest energies. We can conclude that neglecting the correlation result and assuming that all VCV sources 
contribute to the flux the best fit value of the $z_0$ parameter is around $0.06$.

\section{Summary and Conclusions}
\label{sec:conclu}

In this paper we have studied the effects of a local distribution of sources on the UHECR spectrum. In the 
framework of the dip model we have considered a pure proton injection at the source distinguishing among 
two populations of sources: homogeneously distributed at $z>z_0$ and local point sources at lower 
redshift. Following the evidence of correlation published by the Auger collaboration  
\cite{AugerCorrSc,AugerCorrApp,Auger_corr_up} we have considered the VCV catalog of quasars and 
active galactic nuclei as point sources. 

We have shown that under certain assumptions the local sources can induce observable features in the energy
spectrum. In the framework of our model the parameters that play a leading role in fitting the experimental flux are the maximum attainable energy of protons at the source $E_{max}$ and the turning redshift $z_0$. 
We have considered two different scenarios. In the first scenario we have assumed the correlation 
observed by Auger and, according to the Auger analysis, we have fixed $z_0=0.017$ taking 
as local point sources only the $17$ sources of the VCV catalog that show a correlation signal. In the 
second scenario we have neglected the correlation result leaving the parameter $z_0$ free and assuming 
that the contributing sources inside $z_0$ is the complete set of VCV sources. 
 
In the first approach, taking for grant the correlation signal claimed by Auger, our best scenario corresponds to a
choice of $E_{max}$ that should be larger for the local sources respect to the maximum energy associated to the
homogeneously distributed remote sources. Namely $E_{max}=8\times10^{19}$ eV for homogeneous sources
and $E_{max}=10^{21}$ eV in the case of the local sources. The choice of the other injection
parameters, namely $\gamma_g=2.2$ and $m=5$, was fixed comparing the Auger data
\cite{icrc09Schu,SpectrumPLB} with theoretical spectra at energies larger than $3$ EeV. With our best choice for
$E_{max}$ the two-source model explains the small excess at about  $8 \times10^{19}$ eV in the Auger data as 
due to the local sources. In this approach one obtains a very good fit of the experimental data paying the price of
different maximum energies for the two source populations. This assumption is difficult to justify on astrophysical 
grounds.

In order to fix the same maximum energy for all sources, we have considered a second approach. Neglecting 
the Auger result on correlations, we have studied different choices for the turning redshift $z_0$ fixing the same 
injection parameters for all sort of sources, namely $\gamma_g=2.2$, $m=5$ and maximum energy
$E_{max}=3\times 10^{20}$ eV. In this case the best choice for $z_0$ is $z_0=0.06$ and the observed spectrum 
can be fitted up to the highest energies in virtue of the presence of many (around $60$) sources at very low redshift
($z<0.005$) in the VCV catalog. Also in this case the two-source model explains the small excess in the experimental
spectra at about $8\times 10^{19}$ eV as a feature connected with the local distribution of sources.  

It has however to be noted that our model over-fits the experimental data on flux, therefore the $\chi^2$ analysis
cannot exclude a-priori other possibilities, as the case of a homogeneous distribution of sources at any red-shift
\cite{dip}. In the present analysis we have not considered the possibility of nuclei dominated spectra at the highest
energies. This case, favored by the observations of the Auger collaboration on the elongation rate 
\cite{Auger_Xmax}, could be hardly consistent with the correlation result. Nevertheless, the possible correlation 
signal obtained with a mixture of protons and nuclei at the highest energies deserves a more detailed analysis. 

\section*{Acknowledgements}
The authors are very thankful to the Auger group of L'Aquila University: A. Grillo, C. Macolino, 
S. Petrera and F. Salamida, for joint work on this subject. The authors are also grateful to V. Berezinsky
for valuable discussions and a critical reading of the manuscript. DB acknowledges her membership 
in the Pierre Auger Collaboration. This work was partially supported by the contract ASI/INAF I/088/06/0 for
theoretical studies in High Energy Astrophysics and by the Gran Sasso Center for Astroparticle Physics (CFA)
funded by European Union and Regione Abruzzo under the contract P.O. FSE
Abruzzo 2007-2013, Ob. CRO.

\end{document}